\documentclass[11pt]{article}
\usepackage{amssymb}
\usepackage{axodraw}
\usepackage{amsmath}
\usepackage{graphicx}

\newcommand{\ii}{\'\i}
\newcommand{\iii}{\^\i}
\begin{document}

\hfill{}
\begin{tabular}{l}
UB-ECM-PF-01/12 \\
December 2001
\end{tabular}

\begin{center}
{\large \textsc{The scaling evolution of the cosmological constant}} \vskip%
8mm %%%%%%%%%%%%%%%%%%%%%%%%%%%%%%%%%%%%%%%%%%%%%%%%%%%%%%%%%
\textbf{Ilya L. Shapiro}$^{\,a}$\footnote{%
On leave from Tomsk State Pedagogical University, Tomsk, Russia} $\,,\,\,$
\textbf{Joan Sol\`{a}}$^{\,b}$\footnote{%
New permanent address. E-mail: sola@ecm.ub.es. Formerly at the Universitat
Aut\`{o}noma de Barcelona.} \vskip5mm \medskip $^{a}$ \textsl{Departamento
de F\ii sica Te\'{o}rica, Universidad de Zaragoza, 50009, Zaragoza, Spain %
\vskip1mm and Departamento de F\ii sica, ICE, Universidade
Federal de Juiz
de Fora, MG, Brazil} \vskip3mm \medskip %$^{b}$\overleftarrow{}
\textsl{$^{b}$\thinspace\ \thinspace Departament d'Estructura i Constituents
de la Mat\`{e}ria\thinspace }\\[0pt]
\textsl{and Institut de F\ii sica d'Altes Energies, Universitat de
Barcelona,}\\[0pt]
\textsl{Diagonal 647, E-08028, Barcelona, Catalonia, Spain}
\end{center}

\vspace{0.8 cm}

\begin{center}
{ABSTRACT}
\end{center}

\begin{quotation}
In quantum field theory the parameters of the vacuum action are subject to
renormalization group running. In particular, the ``cosmological constant''
is not a constant in a quantum field theory context, still less should be
zero. In this paper we continue with previous work, and derive the particle
contributions to the running of the cosmological and gravitational constants
in the framework of the Standard Model in curved space-time. At higher
energies the calculation is performed in a sharp cut off approximation.
  We assess, in two different frameworks, whether the scaling
dependences of the cosmological and gravitational constants spoil primordial
nucleosynthesis. Finally, the cosmological implications of the running of
the cosmological constant are discussed.
\end{quotation}

\vskip 8mm %%%%%%%%%%%%%%%%%%%%%%%%%%%%%%%%%%%%%%%%%%%%%%%%%%%%%%%%%%%
%% \newpage

\section{Introduction}

The cosmological constant (CC) problem \cite{weinRMP} is, nowadays, one of
the main points of attention of theoretical physics and astrophysics. The
main reason for this is twofold: i) The recent measurements of the
cosmological parameters \cite{Peebles} from high-redshift supernovae \cite
{Supernovae} and the precise data on the temperature anisotropies in the
cosmic microwave background radiation (CMBR) \cite{CMBR,Silk1} offer
unprecedentedly new experimental information on the model universe\,
\footnote{
For a short review of the FLRW cosmological models with non-vanishing
cosmological term, in the light of the recent observations, see e.g. \cite
{SantFeliu}.}; ii) A deeper understanding or even the final solution of the
CC problem is one of the few things that theoretical physics can expect from
the highly mathematized developments of the last decades: from strings and
dualities to the semi-phenomenological Randall-Sundrum model and
modifications thereof \cite{wittenDM,ransum,grs}. The very optimistic
expectation includes also the prediction of the observable particle spectrum
of the Standard Model. However, all attempts to deduce the small value of
the cosmological constant from some sound theoretical idea, without fine
tuning, failed so far and the anthropic considerations could eventually
become a useful alternative to the formal solution from the first principles
of field theory \cite{weinRMP,weinDM}.

In the present paper we continue earlier work on the scaling behavior of the
CC presented in \cite{cosm}. We look at the CC problem using the
Renormalization Group (RG) and the well established formalism of quantum
field theory in curved space-time (see, for example, \cite{birdav,book}).
This way, certainly, does not provide the fundamental solution of the
cosmological constant problem either. Nevertheless it helps in better
understanding the problem and (maybe even more important) in drawing some
physical consequences out of it. The CC problem arises in the Standard Model
(SM) of the strong and electroweak interactions due to the spontaneous
symmetry breaking (SSB) of the electroweak gauge symmetry and the presence
of the non-perturbative QCD vacuum condensates. Both effects contribute to
the vacuum energy density, and when the SM is coupled to classical gravity
they go over to the so-called induced cosmological term, $\Lambda _{ind}$.
The main induced contribution is the electroweak one, roughly given by $%
\Lambda _{ind}\sim M_{F}^{4}$ -- the fourth power of the Fermi scale $%
M_{F}\equiv G_{F}^{-1/2}$. Then the physical (observable) value of the CC
(denoted by $\Lambda _{ph}$) is the sum of the original vacuum cosmological
term in Einstein equations, $\Lambda _{vac}$, and the total induced
contribution, $\Lambda _{ind}$, both of which are individually unobservable.
The CC problem manifests itself in the necessity of the unnaturally exact
fine tuning of the original $\Lambda _{vac}$ that has to cancel the induced
counterpart within a precision (in the SM) of one part in $10^{55}$. These
two: induced and vacuum CC's, satisfy independent renormalization group
equations (RGE). Then, due to the quantum effects of the massive particles,
the physical value of the CC evolves with the energy scale $\mu $: $\Lambda
_{ph}\rightarrow \Lambda _{ph}(\mu )$. Remarkably, the running of the
observable CC has an acceptable range, thanks to the cancellation of the
leading contributions to the $\beta $-functions, which occurs automatically
in the SM \cite{cosm}.

Here we are going to develop the same ideas further. The organization of the
paper is as follows. In the next section, we review the renormalization of
the vacuum action and show that there are no grounds to expect zero CC in
this framework. In section 3, we clarify the source of the cancellation in
the renormalization group equation for the induced CC in the SM. In section
4, we evaluate the value of the CC for higher energies, up to the
electroweak (Fermi) scale and discuss the possible effect of the heavy
degrees of freedom. After that, in section 5, it is verified
whether the running of the CC spoils primordial nucleosynthesis in two
possible frameworks. In section 6 we consider the scaling dependence of the
Newton constant, and show that such dependence cannot be relevant even at
the inflationary scales. In section 7 the role of the CC in the
anomaly-induced inflation is discussed. Finally, in the last section we draw
our conclusions.

%%%%%%%%%%%%%%%%%%%%    ************   %%%%%%%%%%%%%%%%%%%%%
%%%%%%%%%%%%%%%%%%%%    ************   %%%%%%%%%%%%%%%%%%%%%

\section{Renormalization of the vacuum action}

Since we are going to discuss the SM in relation to gravity, it is necessary
to formulate the theory on the classical curved background. In order to
construct a renormalizable gauge theory in an external gravitational field
one has to start from the classical action which consists of three different
parts \footnote{%
See e.g. \cite{birdav,book} for an introduction to renormalization in curved
space-time.}
\begin{equation}
S=S_{m}+S_{nonmin}+S_{vac}\,.  \label{5}
\end{equation}
Here $S_{m}$ is the matter action resulting from the corresponding action of
the theory in flat space-time after replacing the partial derivatives by the
covariant ones, Minkowski metric by the general metric and the integration
volume element $d^{4}x$ by $d^{4}x\sqrt{-g}$. For instance, the scalar
kinetic term and the Higgs potential enter (\ref{5}) through
\begin{equation}
S_{sc}=\int d^{4}x\sqrt{-g}\,\left\{ g^{\mu \nu }\,\left( D_{\mu }\Phi
\right) ^{+}\left( D_{\nu }\Phi \right) -V_{0}(\Phi )\right\} \,,  \label{Sc}
\end{equation}
where the derivative $D_{\mu }$ is covariant with respect to general
coordinate transformations and also with respect to the gauge
transformations of the SM electroweak symmetry group $SU(2)_{L}\times
U(1)_{Y}$. Thus
\begin{equation}
D_{\mu } =\nabla _{\mu }-ig\,T^{i}\,W_{\mu }^{i}-ig^{\prime }\,{Y}B_{\mu },
\label{Cov}
\end{equation}
where $\nabla _{\mu }$ is the coordinate-covariant part and the rest of the
terms involve the standard gauge connections formed out of the electroweak
bosons $W_{\mu }^{i},\,B_{\mu }$ and the corresponding gauge couplings and
generators. Other terms in the action involve similar generalization of the
ordinary fermion, gauge and Yukawa coupling interactions of the SM. One of
the novel features of the SM in a curved space-time background is the
necessity of the ``non-minimal term''
\begin{equation}
S_{nonmin}=\int d^{4}x\sqrt{-g}\,\xi \,\,\Phi ^{\dagger }\Phi \,R,  \label{6}
\end{equation}
involving the interaction of the $SU(2)_{L}$ doublet of complex scalar
fields $\Phi $ with the curvature scalar $R$. Notice that for $\,\xi =0$ the
gravitational field is still (minimally) coupled to matter through the
metric tensor in the kinetic terms and in general to all terms of the
Lagrangian density through the $\sqrt{-g}$ insertion. With respect to the
RG, one meets an effective running of $\xi $ whose value depends on scale.
This running has been studied for a variety of models (see \cite{tmf,book}
and references therein).

Very important for our future considerations are the vacuum terms in the
action, which are also required to insure renormalizability and hence repeat
the form of the possible counterterms:
\begin{equation}
S_{vac}=\int d^{4}x\sqrt{-g}\,\Big\{ \,a_{1}R^{2}_{\mu\nu\alpha\beta}
+a_{2}R^{2}_{\mu\nu} + a_{3}R^{2} + a_{4}{\square} R-\frac{1}{16\pi G_{vac}}%
\,R - \Lambda_{vac}\Big\}  \label{7}
\end{equation}

All the divergences in the theory (\ref{5}) can be removed by the
renormalization of the matter fields and couplings, masses, non-minimal
parameter $\xi$ and the bare parameters of the vacuum action $a_{1,2,3,4},$ $%
\,G_{vac}\,$ and $\,\Lambda_{vac}$.

Our main attention will be paid to the cosmological term in the vacuum
action (\ref{7}). Formally, the vacuum CC is required for renormalizability
even in flat space-time. But then it is just a constant addition to the
Lagrangian, which does not affect the equations of motion. In curved
space-time, however, the situation is quite different, because the CC
interacts with the metric through the $\sqrt{-g}$ insertion. This was first
noticed by Zeldovich \cite{zeldo}, who also pointed out that there is a
cosmological constant induced by matter fields.

Let us briefly describe the divergences in the vacuum sector. Suppose that
one applies some covariant regularization depending on a massive parameter.
For example, it can be regularization with higher derivatives \cite
{slavnov,asorey} with additional Pauli-Villars regularization in the
one-loop sector. To fix ideas let us assume that all of the divergences
depend on a unique regularization parameter $\Omega$ with dimension of mass.
Then, for the renormalizable theory (\ref{7}), one faces three types of
divergences in the vacuum sector: \vskip2mm

i) Quartic divergences $\,\,\sim \,\Omega ^{4}\,\,$ for the cosmological
term come from any field: massive or massless. As usual, these divergences
must be subtracted by a counterterm. The renormalization condition for the
cosmological constant can be fixed at some scale (see below) in such a way
that there should not be running due to these divergences. They fully cancel
out against the counterterm; in other words, they can be ``technically''
disposed of once and forever. That is why the vacuum oscillations (zero-mode
contributions) of the fields \cite{weinRMP}, although they could reach
values as large as the Planck mass $M_{P}$ to the fourth power, do not pose
a severe problem and are usually considered as unimportant .

ii) Quadratic divergences are met in the Hilbert-Einstein and CC terms $%
\,\,1/(16\pi G_{vac})\,R+\Lambda _{vac}\,\,$ of Eq. (\ref{7}). They can
arise either from quadratically divergent graphs or appear as sub-leading
divergences of quartic divergences. In the CC sector they are proportional
to $\,\Omega ^{2}m_{i}^{2}$, where $m_{i}$ are masses of the matter fields.
Hence, massless fields do not contribute to these divergences. Quadratic
divergences are removed in the same manner as the quartic ones. One can
always construct the renormalization scheme in which no quadratic scale
dependence remains after the divergences were canceled and the
renormalization condition fixed.

iii) Finally, there are the logarithmic divergences. They show up in all
sectors of Eq. (\ref{7}), and come from all the fields: massless and
massive. However, only massive fields contribute to the divergences of the
CC and Hilbert-Einstein terms. This can be easily seen from dimensional
analysis already. The logarithmic divergences are the most complicated ones,
because even after been canceled by counterterms their effect is spread
through the renormalization group.

What is the experimental situation at present?
By virtue of the recent astronomical observations from high
red-shift supernovae \thinspace \cite{Supernovae}, the
present-day density of matter and the CC are given by $\rho
_{M}^{0}\simeq 0.3\,\rho _{c}^{0}$ and $\Lambda _{ph}\simeq
0.7\,\rho _{c}^{0}$, respectively, where
\begin{equation}
\rho _{c}^{0}\equiv 3\,H_{0}^{2}/8\,\pi \,G_{N}\simeq
8.1\,h_{0}^{2}\times 10^{-47}\,GeV^{2}=\left(
3.0\,\sqrt{h_{0}}\times 10^{-3}\,eV\right) ^{4} \label{rhoc0}
\end{equation}
is the critical density \thinspace \cite{Peebles}. Here the dimensionless
number $h_{0}=0.65\pm 0.1$ \cite{CMBR,Silk1} defines the experimental range
for the present day Hubble's constant
\begin{equation}
H_{0}\equiv 100\,h_{0}\ \,Km\,sec^{-1}\,Mpc^{-1}\simeq 2.13\,h_{0}\times
10^{-42}\,GeV\,.  \label{H0}
\end{equation}
The next step is to choose the renormalization condition at some fixed
energy scale. The choice of the scale $\mu $ is especially relevant for the
CC, because the latter is observed only at long, cosmic distances,
equivalently at very low energies. We shall identify the meaning of $\mu $
more properly later on, but for the moment \ it suffices to say that our
considerations on the cosmological parameters always refer to some specific
renormalization point, which we denote $\mu _{c}$. The choice of $\mu _{c}$
must be such that at lower energies, $\mu <\mu _{c}$, there is no running.
From the RG \ analysis we may expect\thinspace \cite{cosm}
$\Lambda _{ph}$ $\sim $ $\mu _{c}^{4}$ \ and so from the above experimental
results we must have
\begin{equation}
\mu _{c}=\mathcal{O}(10^{-3})\,eV\,.  \label{muc}
\end{equation}
Since the renormalization of the CC is only due to the
contributions of massive particles, one may guess that $\mu _{c}$ must be of
the order of the lightest particle with non-vanishing mass. \ In a minimal
extension of the SM, it is the lightest neutrino (denoted here as $\nu _{1}$%
), so\thinspace\ we may expect \ $\mu _{c}\approx m_{\nu _{1}}=\mathcal{O}%
(10^{-3})\,eV\,\,$, which in fact holds good if we rely on the
current results on neutrino masses \cite{Valle}. \ Whether this
is a coincidence or not cannot be decided at this stage, but one
can offer RG arguments in favor of it\ \cite{cosm}.
Actually,
from these results we collect, not one but three ``cosmic
coincidences''; i) The physical value of the CC is positive and
of the order of $\rho _{M}^{0}$; ii) The density of matter is of
the order of the critical density ($\rho _{M}^{0}\sim \rho
_{c}^{0}$); and iii) $\ m_{\nu
_{1}}^{4}$ is of the order of $\rho _{M}^{0}$ and so of order $\Lambda _{ph}$%
. Hence \footnote{%
The contribution to $\rho _{M}^{0}$ from light neutrinos is\thinspace \cite
{Peebles} $\rho _{M}^{(\nu )}\,h_{0}^{2}\simeq \left( m_{\nu
_{1}}/92\,eV\right) \rho _{c}^{0}$. In view of eq.(\ref{muc}), $\rho
_{M}^{(\nu )}$ is much smaller than $m_{\nu _{1}}^{4}$. Heavier neutrino
species also contribute, but it is not clear why the full matter density
happens to coincide (in order of magnitude) with the fourth power of the
mass of the lightest species.}
\begin{equation}
\left( \Lambda _{ph}\right) ^{1/4}\sim \left( \rho _{c}^{0}\right)
^{1/4}=3.0\,\sqrt{h_{0}}\times 10^{-3}\,eV\sim m_{\nu _{1}}  \label{muc2}
\end{equation}
As we shall see in the following, these coincidences are
perhaps not independent, if the CC problem is addressed from the
RG point of view.

At energies up to the Fermi scale, $\mu \lesssim M_{F}$, the higher
derivative terms in (\ref{7}) are not important for our considerations, and
so the renormalized effective vacuum action at these low energies is just
the Hilbert-Einstein action with a running cosmological and gravitational
constants $G_{vac}(\mu ),\,\,\Lambda _{vac}(\mu )$:
\begin{equation}
S_{HE}(\mu )=-\int d^{4}x\sqrt{-g}\,\left\{ \,\frac{1}{16\pi G_{vac}(\mu )}%
\,R+\Lambda _{vac}\,(\mu )\,\right\} \,.  \label{HE}
\end{equation}

In the last equation the dependence on $\mu $ is governed by the
renormalization group. At lower energies heavy particles decouple and do
not, in principle, contribute to the running. However, this consideration
must be handled with care in the CC framework. For a better understanding of
the decoupling of heavy particles we have to remember that the decoupling
mechanism \cite{AC} compares the mass of the particle inside the quantum
loop with the energy of the particles in the external lines connected to
this loop. Here, we are dealing with the vacuum diagrams
and the gravitational field and therefore the relevant external legs to
consider are the ones of gravitons\thinspace \footnote{%
The term graviton is used here in a generic sense referring to the presumed
quantum of gravity as a field theory with a tensor potential, rather than to
the gravitational waves.}. The next problem is to evaluate the energy of
these gravitons. Indeed, there is no general method to estimate the energy
of the gravitational field, so we have some freedom at this point. One of
the possibilities is the following. At low energy the dynamics of gravity is
defined by the Einstein equations
\begin{equation}
R_{\mu \nu }-\frac{1}{2}\,Rg_{\mu \nu }=8\pi G_{ph}\,(T_{\mu \nu }+g_{\mu
\nu }\,\Lambda _{ph}\,)\,,  \label{Einstein}
\end{equation}
where $G_{ph}=1/M_{P}^{2}$ is the physical value of Newton's constant--see
section 6. We may use the value of the curvature scalar ($%
R$) as an order parameter for the gravitational energy, so the RG
scale $\mu $ can be associated with $R^{1/2}$. From
eq.(\ref{Einstein}) \ we see that this is equivalent to take $\mu
\sim $ $\sqrt{T_{\mu }^{\mu }/M_{P}^{2}}$. But in the
cosmological setting the basic dynamical equations refer to the
scale factor $a(t)$ of the Friedmann-Lema\iii tre-Robertson-Walker
(FLRW) metric\thinspace \cite{Peebles}, and so we must re-express
the graviton energy in terms of it. The $00$ component of
(\ref{Einstein}) yields the well-known Friedmann-Lema\iii tre
equation
\begin{equation}
H^{2}\equiv \left( \frac{\dot{a}}{a}\right) ^{2}=\frac{8\pi }{3\,M_{P}^{2}}%
\left( \rho +\Lambda _{ph}\right) -\frac{k}{a^{2}}\,.  \label{FL}
\end{equation}
The space curvature term can be safely set to zero ($k=0$),
because the universe is very flat at present\thinspace \cite{Peebles} and in
general it may have undergone an inflationary period \cite{Guth}\thinspace
,\ so that $k=0$ \ effectively throughout the whole FLRW regime. The spatial
components of (\ref{Einstein}), combined with the $00$ component (\ref{FL}),
yields the following dynamical equation for $a(t)$:
\begin{equation}
\ddot{a}=-\frac{4\pi }{3\,M_{P}^{2}}\left( \rho +3\,p-2\,\Lambda
_{ph}\right) \,a\,.  \label{accel}
\end{equation}
In these equations $\rho =\rho _{M}+\rho _{R}$ is the total energy density
of matter and radiation, and $p$ is the pressure. In the modern Universe $%
p\simeq 0$ and $\rho \simeq \rho _{M}^{0}$. Moreover, from the recent
supernovae data\thinspace \cite{Supernovae}, we know that $\Lambda _{ph}$
and $\rho _{M}^{0}$ have the same order of magnitude as the critical density
$\rho _{c}^{0}$. Therefore, the source term on the \textit{r.h.s.} of (\ref
{accel}) is characterized by a single dimensional parameter $\sqrt{\rho
_{c}^{0}/M_{P}^{2}}$, which according to eq. (\ref{FL}) is nothing but the
experimentally measurable Hubble's constant $H_{0}$. This is obviously
consistent with the expected result $\sqrt{T_{\mu }^{\mu }/M_{P}^{2}}$ in
the general case because $T_{\mu }^{\mu }\sim \rho _{M}^{0}\sim \rho _{c}^{0}
$ for the present-day universe. \ Therefore, we conclude that
\begin{equation}
\mu \sim R^{1/2}\sim H(t)\,,  \label{muH}
\end{equation}
can be the proper identification for the RG scale $\mu $ in the
semiclassical treatment of the FLRW cosmological framework. Another possible
choice (the one advocated in \cite{cosm}) is to consider the critical
density,
\begin{equation}
\,\mu \sim \rho _{c}^{1/4}(t)\,,
\label{muCD}
\end{equation}
as a typical energy of the Universe. During the hot
stages of its evolution (see Section\ 5) this entails a direct association
with temperature $\mu \sim T$ \cite{cosm,Babic}. Although there might be
other reasonable possibilities we will only consider these two $\ $``$\mu $%
-frames'' in the present paper. In the following we will evaluate the
running of the CC and Newton's constant with respect to the change in
magnitude of the graviton energy and only after that we will discuss the
scale at which these estimates might be applied and the differences between
the two choices of $\mu $.

%%%%%%%%%%%%%%%%%%%%%%%%%%%%%%%%%%%%%%%%%%%%%%%%
Let us postpone the discussion of the renormalization of the other terms in (%
\ref{7}) until section 7 and concentrate now on the CC. As stated, the value
of the CC is supposed to be essentially constant between $\mu _{c}$ and the
present cosmic scale $H_{0}$. Hence, one can impose the renormalization
condition at $\mu _{c}$. As soon as one deals with the SM in curved
space-time, the vacuum parameters including $\Lambda _{vac}$, should be
included into the list of parameters of the SM. These parameters must be
renormalized and their physical values should be implemented via the
renormalization conditions. Exactly as for any other parameter of the SM,
the values of $\Lambda _{vac}$ and $\xi ,G_{vac},a_{1,2,3,4}$ should result
from the experiment. However, there is an essential difference between the
CC and, say, masses of the particles. The key point of the CC problem is
that there is another ''induced'' contribution $\,\,\Lambda _{ind}\,\,$ to
the CC along with the $\Lambda _{vac}$. The observable CC is the sum of the
vacuum and induced terms
\begin{equation}
\Lambda _{ph}=\Lambda _{vac}+\Lambda _{ind}\,,  \label{lambda_phys}
\end{equation}
evaluated at the cosmic renormalization scale $\mu _{c}$.

The values of the electroweak parameters of the SM are defined from high
energy experiments. The characteristic scale in this case is the Fermi scale
$M_{F}\equiv G_{F}^{-1/2}\simeq 293\,GeV$. At the same time, due to the
weakness of the gravitational force at small distances, there is no way to
measure the vacuum parameters at this scale. However, recent astronomical
observations are currently being interpreted as providing the right order of
magnitude of the ``physical'' cosmological constant at present, $\Lambda
_{ph}$, and it comes out to be non-zero at the $99\%$ C.L. \cite{Supernovae}%
. Now, the value of $\Lambda _{ph}$ derived from these observations will be
treated here as the value of the running parameter $\Lambda _{ph}(\mu )$
evaluated at $\mu =\mu _{c}$.

Let us now review the mechanism for the induced CC, in the electroweak
sector of the SM. In the ground (vacuum) state of the SM, the expectation
value (VEV) of $\,\Phi ^{+}\Phi \,$ in (\ref{Sc}) will be denoted $\,<\Phi
^{+}\Phi >\equiv \frac{1}{2}\phi ^{2}$, where $\,\phi \,$ is a classical
scalar field. The corresponding classical potential reads
\begin{equation}
V_{cl}=-\frac{1}{2}m^{2}\phi ^{2}+\frac{f}{8}\phi ^{4}.  \label{2a}
\end{equation}
Shifting the original field $\phi \rightarrow H^{0}+v$ such that the
physical scalar field $H^{0}$ has zero VEV one obtains the physical mass of
the Higgs boson: $M_{H}=\sqrt{2}\;m$. Minimization of the potential (\ref{2a}%
) yields the SSB relation:
\begin{equation}
\phi =\sqrt{\frac{2m^{2}}{f}}=v\,\,\,\,\,\,\mathrm{and} \,\,\,\,\,\,\,f=%
\frac{M_{H}^{2}}{v^{2}}\,.  \label{5N}
\end{equation}
The VEV $<\Phi >\equiv v/\sqrt{2}$ gives masses to fermions and weak gauge
bosons through
\begin{equation}
m_{i}=h_{i}\,\frac{v}{\sqrt{2}}\,,\;\;\;M_{W}^{2}=\frac{1}{4}
\,g^{2}\,v^{2},\;\;\;M_{Z}^{2} =\frac{1}{4}\,(g^{2}\,+g^{\prime 2})\,v^{2},
\label{masses}
\end{equation}
where $h_{i}$ are the corresponding Yukawa couplings, and $g\,$ and $%
\,g^{\prime }\,$ are the $SU(2)_{L}$ and $U(1)_{Y}$ gauge couplings. The VEV
can be written entirely in terms of the Fermi scale: $\,v=2^{-1/4}M_{F}%
\simeq 246\,GeV$. From (\ref{5N}) one obtains the following value for the
potential, at the tree-level, that goes over to the induced CC:
\begin{equation}
\Lambda _{ind}=<V_{cl}>=-\frac{m^{4}}{2f}\,.  \label{nnn6}
\end{equation}
If we apply the current numerical bound $\,M_{H}\gtrsim 115\,GeV$ from LEP
II, then the corresponding value $\left| \Lambda _{ind}\right| \simeq
1.0\times 10^{8}\,GeV^{4}$ is $55$ orders of magnitude greater than the
observed upper bound for the CC -- typically this bound is $\Lambda
_{ph}\lesssim 10^{-47}GeV^{4}$.

In order to keep the quantum field theory consistent with astronomical
observations, one has to demand that the two parts should cancel with the
accuracy dictated by the current data. This defines the sum (\ref
{lambda_phys}). As shown by Eq.(\ref{nnn6}), the first term $\Lambda _{ind}$
on the \textit{r.h.s.} of (\ref{lambda_phys}) is not an independent
parameter of the SM, since it is constructed from other parameters like the
VEV of Higgs and couplings. On the contrary, $\Lambda _{vac}$ is an
independent parameter and requires an independent renormalization condition.
From the quantum field theory point of view, the sequence of steps in
defining the CC is the following: one has to calculate the value of $\Lambda
_{ind}$ at $\mu _{c}$, measure the value of the physical CC, $\Lambda _{ph}$%
, at the same scale, and choose the renormalization condition for $\Lambda
_{vac}$ in the form
\begin{equation}
\Lambda _{vac}({\mu _{c})}=\Lambda _{ph}({\mu _{c}})-\Lambda _{ind}({\mu _{c}%
})\,.  \label{ren_cond}
\end{equation}
The modern observations from the supernovae \cite{Supernovae} tell us that
the value of $\Lambda _{ph}(\mu =\mu _{c})$ is positive and has the
magnitude of the order of $\rho _{c}^{0}$ , that is about $10^{-47}\,GeV^{4}$%
. This value should be inserted into the renormalization condition (\ref
{ren_cond}). From the formal point of view, everything is consistent. There
is no reason to insist that the CC should be exactly zero, for it is
measured to be nonzero by experiment. In principle, since the
renormalization condition for the CC should be taken from the measurement,
to insist on any other value, including zero, is senseless. But, the problem
with the Eq. (\ref{lambda_phys}) is that the terms on the \textit{r.h.s.} of
it are 55 orders greater than their sum, so that one has to define $\Lambda
_{vac}({\mu _{c})}$ with the precision of 55 decimal orders. To explain this
fantastic exactness is the CC problem. Of course, the fine tuning of 55
decimal numbers is difficult to understand, but zero CC would mean
infinitely exact fine tuning, which would be infinitely hard to explain, at
least from the RG point of view. Let us compare the CC with any other
parameter of the SM. Imagine, for instance, that we could isolate some
particle like the electron or the top quark. Suppose also that we could
measure its mass with the $55^{th}$ order of magnitude precision. Then we
meet a similar problem, because we would not be able to explain why the mass
of this particle is exactly that one we measure.
%% As a matter of fact, we are not able to explain
%% the values of these masses even when taken with their present accuracy.
Indeed, for the electron (not to mention the top quark!) the $55^{th}$ order
of magnitude precision is not possible, so the exactness of the ``fine
tuning'' for the CC really looks as something outstanding. However, this
just manifests the fact that the cosmic scale where the measurement is
performed, is quite different from the Fermi scale, where the second
counterpart $\Lambda _{ind}$ is defined. Therefore, the difference between
the CC and the particle masses is that the first one can only be measured at
the cosmic scale. Unfortunately, the problem of CC is deeper than that. Let
us continue our comparison with the electron mass. It is known to be
$\,m_{e}=0.51099906(15)\,MeV$. But if it would be, say, $\,m_{e}=0.52\,MeV$,
physics should be, perhaps, the same. At the same time, if we change, for
example, the last 7 decimal points in the 55-digit number for the modern
value of the CC, the energy density of the Universe would change a lot and
the shape of the whole Universe would look very different. For instance, the
Universe could be in a state of fast inflation. Thus, the problem of the
fine tuning of the CC is much more severe than the prediction of particle
masses.

The point is that we do not know why our Universe, with its small value of
the CC, is what it is. This can be taken as a philosophic question, but if
taken as a physical problem, it is really difficult to solve. At present,
there are three main approaches. One of them supposes that there is some
hidden symmetry which makes CC exactly zero, for instance, at the zero
energy scale \cite{wittenDM} %%%%%%%%%%%%%%%%%%%%%%%%%%%%% FOOTNOTE
\footnote{%
One can, also, mention the interesting recent papers \cite{Volovik} where
the methods of condensed matter physics were called to solve the CC problem.
According to \cite{Volovik}, the CC$=0$ in the infinitely remote future (for
the open Universe) follows from the condition of equilibrium for the matter
filling this Universe. While being a useful observation, this does not solve
the fundamental CC problem. Indeed, the presence of the CC causes the matter
filling the Universe to be in a non-equilibrium state in the remote future.
Thus, to postulate as a fundamental principle that the the matter
\textit{should} eventually reach the state of equilibrium is just a version
of the
standard fine tuning of the CC.}. Then one has to explain the change to the
nonzero value of the CC at the present cosmic scale. The second possibility
is the anthropic hypothesis, which supposes that our Universe is such as it
is because we are able to live in it and study it. An extended version
supposes multiple universes and challenges one to calculate the probability
to meet an appropriate Universe, available for doing theoretical physics
\cite{antrop}. \ The third approach is called quintessence and assumes that
the CC is nothing but a scalar field providing negative pressure and a
time-varying, spatially fluctuating energy density \cite{Caldwell}.

Let us consider the scale dependence in the RG framework. When the Universe
evolves, the energy scale changes and it is accompanied by the scale
dependence (running) of the physical quantities like charges and masses.
Taking the quantum effects into account, one cannot fix the CC to be much
smaller than $\,10^{-47}\,GeV^{4}$, because such a constraint would be
broken by the RG at the energies comparable to the small neutrino masses
\cite{cosm}. In fact, we can accept (\ref{ren_cond}) as an experimental
fact, without looking for its fundamental reasons. Then, the following
questions appear: i) Is the running of the CC consistent with the standard
cosmological model?; $\,\,$ ii) Which kind of lessons can we learn from this
running?.

As we have shown in \cite{cosm} the running for the observable CC really
takes place. The RGE for the parameter $\Lambda _{vac}$ is independent from
the RG behavior of the induced value $\Lambda _{ind}$, and as a result the
sum (\ref{lambda_phys}) diverges from its value at the fixed cosmic scale
(\ref{ren_cond}). It is important to notice that the running of the physical
(observable) CC signifies that one cannot have zero CC during the whole life
of the Universe because a CC of the order of the $\beta $-function would
appear at the neutrino scale. For this reason, popular quintessence models,
which are called to mimic the CC cannot ''explain'' the observed value of
the CC \cite{weinDM}. With or without quintessence, one has to choose the
renormalization condition for the vacuum CC. The only difference is that, in
(\ref{ren_cond}), one has to add the quintessence contribution to $\,\Lambda
_{ind}$, and the fine tuning becomes a bit more complicated. Hence,
quintessence can be useful to solve some other problem, but in our opinion
it does not simplify at all the CC problem.

%%%%%%%%%%%%%%%%%%%%    ************   %%%%%%%%%%%%%%%%%%%%%
%%%%%%%%%%%%%%%%%%%%    ************   %%%%%%%%%%%%%%%%%%%%%
%%%%%%

\section{Running of vacuum and induced counterparts}

%%%%%%%%%%%%%%%%%%%%    ************   %%%%%%%%%%%%%%%%%%%%%

The RGE for the effective action can be formulated in curved space-time in
the usual form \cite{tmf,book}:
\begin{equation}
\left\{ \mu \frac{\partial }{\partial \mu }+\left( \beta _{p}-d_{p}p\right)
\,\frac{\partial }{\partial p}+\gamma _{i}\int d^{4}x\sqrt{-g}\;\Phi _{i}\,
\frac{\delta }{\delta \Phi _{i}}\right\} \, \Gamma \left[ g_{\alpha
\beta},p,\Psi _{i},\mu \right] =0\,.  \label{9}
\end{equation}
Here the $\beta $-functions for all the couplings, vacuum parameters and
masses of the theory (generically denoted by $p$) and the $\gamma_i$%
-functions of all the matter fields $\Phi_i$ are defined in the usual way.

Equation (\ref{9}) enables one to investigate the running of the coupling
constants and also the behavior of the effective action in a strong
gravitational field, strong scalar field and other limits \cite{book}. We
are interested in the running (general dependence on $\mu $) of the CC and
Newton's constant. Also, we shall consider the RGE's for other parameters
when necessary.

To study the running of the physical CC and also, in a subsequent section,
for the Newton constant, we need the $\beta $-functions for the scalar
coupling constant $f$, for the Higgs mass parameter $m$ and for the
dimensional parameters $\,G_{vac},\,\Lambda _{vac}\,$ of the vacuum action.
At this stage we write down the full RGE's without restrictions on the
contributions of heavy particles. These restrictions will be imposed later,
when we evaluate the running at different energies. Taking into account all
the fields entering the SM we arrive at the following RGE:
\begin{align}
(4\pi )^{2}\frac{dm^{2}}{dt}& =m^{2}\;\left( 6f-\frac{9}{2}\;g^{2}
-\frac{3}{2}\;g^{\prime }{}^{2}+2\,\sum_{i=q,l}\,N_{i}\,h_{i}^{2}\right)
,\ \ \ \ \ m(0)=m(\mu =M_{F})\equiv m_{F}\,,  \notag \\
(4\pi )^{2}\frac{df}{dt}& =12f^{2}-9fg^{2}-3fg^{\prime }{}^{2} +\frac{9}{4}
g^{4}+\frac{3}{2}g^{2}g^{\prime }{}^{2}+\frac{3}{4}g^{\prime }{}^{4}  \notag
\\
& +4\sum_{i=q,l}N_{i}\,h_{i}^{2}\left( f-h_{i}^{2}\right) \ \ \ \ \ \ \ \ \
\ \ \ \ \ \ \ \ \ \ \ \ \ \ \ \ \ \ \ \ \ \ f(0)=f(\mu =M_{F})\equiv f_{F}\,.
\label{n110}
\end{align}
Here $\,\,h_{i}=h_{l,q}\,\,$ are the Yukawa couplings for the fermions:
quark $\,q=(u,..,t)\,$ and lepton $\,l=(\nu _{e},\nu _{\tau },\nu _{\mu
},e,\mu ,\tau )\,$ constituents of the SM. $\,$ Furthermore, $\,t=\ln (\mu
/M_{F})$, and $\,N_{i}=1,3\,$ for leptons and quarks respectively. The
boundary conditions for the renormalization group flow are imposed at the
Fermi scale $M_{F}$ for all the parameters, with the important exception of
$\Lambda _{vac}$. Then the $SU(2)_{L}$ and $U(1)_{Y}$ gauge couplings at $\mu
=M_{F}$ \ are $\,g_{F}^{2}\approx 0.4\,$ and $\,g^{\prime
}{}_{F}^{2}=\,g_{F}^{2}\tan ^{2}\theta _{W}\,\simeq 0.12.$ Here $\theta _{W}$
is the weak mixing angle, and at the Fermi scale $\sin^{2}\theta _{W}\simeq
0.23$.

Taking the renormalization conditions into account, the solution of (\ref
{n110}) for $m$ can be written in the form:
\begin{equation}
m^{2}(t)=m_{F}^{2}\,U(t),  \label{n112}
\end{equation}
with
\begin{equation}
U(t)=exp\left\{ \int_{0}^{t}\,\frac{dt}{(4\pi )^{2}}\, \left[ 6\,f(t)
-\frac{9}{2}\,g^{2}(t)-\frac{3}{2} \,g^{\prime 2}(t)
+\sum_{i}\,N_{i}\,h_{i}^{2}(t)
\right] \right\} \,\,,  \label{n113}
\end{equation}
where the couplings satisfy their own (well known) RGE \cite{CEL}. The
one-loop $\beta$-function for the vacuum CC gains contributions from all
massive fields, and can be computed in a straightforward way by explicit
evaluation of the vacuum loops (Cf. \textit{Fig.1}) . In particular, the
contribution from the complex Higgs doublet $\Phi $ and the fermions is (for
$\mu \gtrsim M_{F}$)
\begin{equation}
(4\pi )^{2}\frac{d\Lambda _{vac}}{dt}\,=\beta _{\Lambda }\equiv
\,2\,m^{4}\,-\,2\,\sum_{i}\,N_{i}\,m_{i}^{4}
\,,\;\;\;\;\;\;\;\;\;\;\;\;\;\;\;\;\;\;\;\;\;\;\;\;\;\;\;\; \Lambda
_{vac}(0)=\Lambda _{0}\,,  \label{n11}
\end{equation}
where the sum is taken over all the fermions with masses $m_{i}$. \ In the
last formula we have changed the dimensionless scaling variable into
$\,t=\ln (\mu /\mu _{c})$ because, as we have already argued above, the
renormalization point for $\Lambda _{vac}$ is $\mu =\mu _{c}$. Taking (\ref
{n112}) into account, the solution for the vacuum CC is
\begin{equation}
\Lambda _{vac}(t)=\Lambda _{0}+\frac{2\,m_{F}^{4}}{(4\pi )^{2}}
\;\;\int_{0}^{t}U^{2}(t)\,dt-\frac{2}{(4\pi )^{2}}\sum_{i}\,N_{i}
\int_{o}^{t}m_{i}^{4}(t)\,dt\,,  \label{n114}
\end{equation}
where the running of $m_{i}(t)$ is coupled to that of $m^{2}(t)$ and the
corresponding Yukawa couplings. The scale behavior of $\Lambda _{ind}$
depends on the running of $m(t)$ and $f(t)$, so that from Eq.(\ref{nnn6}) we
have
\begin{equation}
\Lambda _{ind}(t)=-\frac{m_{F}^{4}\;U^{2}(t)}{2f(t)},  \label{n117}
\end{equation}
where $f(t)$ is solution of Eq. (\ref{n110}). Although the value of the
Higgs mass is not well under control at present, and therefore the initial
data for $f$ is unknown, this uncertainty does not pose a problem for the
running of the CC, especially at low energies where the heavy degrees of
freedom play an inessential role. Eqs. (\ref{n114}), (\ref{n117}) enable one
to write the general formula for the scale dependence of the CC, in a
one-loop approximation:
\begin{equation}
\Lambda _{ph}(t)=\Lambda _{0}-\frac{m_{F}^{4}\,\,U^{2}(t)}{2f(t)}
+\frac{2\,m_{F}^{4}}{(4\pi )^{2}}\;\int_{0}^{t}U^{2}(t)dt
-\frac{2}{(4\pi )^{2}}
\,\sum_{i}\,N_{i}\int_{o}^{t}m_{i}^{4}(t)\,dt\,,  \label{nn117}
\end{equation}
where $t=\ln (\mu /\mu _{c})$. An important point concerning the RGE is the
energy scale where they actually apply. This is especially important in
dealing with the CC problem, since this problem is seen at the energies far
below the Standard Model scale ($\mu_{c}\ll M_{F}$). The corresponding
$\beta $-functions $\,\,\beta _{\Lambda _{vac}},\beta _{m},\beta
_{h_{i}},\beta_{f}\,...\,$ depend on the number of active degrees of
freedom. These are the number of fields whose associated particles have a
mass below the typical energy scale $\mu$ of the gravitons (e.g., external
legs of the diagrams which must be, indeed, added at \textit{Fig. 1}),
because at sufficiently small energies one can invoke the decoupling of the
heavier degrees of freedom \cite{AC}. Equation (\ref{nn117}) is normalized
according to (\ref{ren_cond}) such that the quantity $\ \Lambda_{ph}(0)$
exactly reproduces the value of the CC from supernovae data. Therefore, it
should be clear that in our framework the relevant CC at present is not the
value of (\ref{n117}) at $\mu =M_{F}$ but, \ instead, that of (\ref{nn117})
at $\mu =\mu _{c}\ll M_{F}$. \ The value of the CC at the Fermi scale will
be computed below within our approach.

First we will be interested in the scaling behavior of $\,\Lambda _{ph}$
starting from the low energy scale $\mu \sim \mu _{c}$. One may expect that
the lightest degrees of freedom of the SM, namely the neutrinos, are the
only ones involved to determine the running $\Lambda _{ph}$ at nearby points
$\mu \gtrsim \mu _{c}$. Thus, let us suppose that all other constituents of
the SM decouple, including the heavier neutrino species (see below) and of
course all other fermions, scalar and gauge bosons. For example, the
electron (which is the next-to-lightest matter particle after the neutrino)
has a mass which is $10^{8\text{ }}$ times heavier than the assumed mass for
the lightest neutrino species \cite{Valle}. Within this Ansatz, we have to
take into account only light neutrino loops. Moreover, we can safely neglect
the running of the mass $m(t)$ and coupling $f(t)$, and attribute their
values at the Fermi scale to them. \ For, the effect of their running at one
loop is of the same order as the second loop corrections to the running of
$\Lambda_{vac}$ and $\Lambda_{ind}$, because they are proportional to the
same neutrino Yukawa couplings.

Substituting (\ref{n110}) into the expression (\ref{nnn6}), we arrive at the
following equation:
\begin{equation}
\frac{d\Lambda _{ind}}{dt}\,=\,\frac{m^{4}}{2f^{2}}\,\frac{df}{dt}\, -\,
\frac{m^{2}}{f}\,\frac{dm^{2}}{dt}\, =\,-\,\frac{1}{(4\pi )^{2}}\cdot
\frac{2m^{4}}{f^{2}}\,\cdot \,\sum_{j}h_{j}^{4}\,=\,-\frac{2}{(4\pi )^{2}}\,
\sum_{j}m_{j}^{4}\;.  \label{ccel1}
\end{equation}
Here we have used the fact that in the SM the coefficient $%
\,\,m^{4}h_{j}^{4}/f^{2}\,\,$ is nothing but $\ $ the fourth power of the
fermion mass, $m_{j}^{4}$ - as it follows from Eqs. (\ref{5N}) and (\ref
{masses}). In fact, the \textit{r.h.s.} of (\ref{ccel1}) looks like a
miracle occurring in the SM, because it exhibits a cancellation of the
leading $\,m^{4}h_{j}^{2}/f\,$ terms. These terms are $28$ orders of
magnitude greater than the remaining ones $\,m^{4}h_{j}^{4}/f^{2}$ in the
case of the lightest neutrinos (as seen by using Eqs. (\ref{5N},\ref{masses}),
the ratio of the two is $f/h^{2}=M_{H}^{2}/2m_{\nu }^{2}\sim 10^{28}$ for
$m_{\nu }\sim 10^{-3}\,eV$). Without this cancellation the range of the
running would be unacceptable, and the fine tuning of the CC incompatible
with the standard cosmological scenario (see section 5). This does not
happen in the SM due to the mentioned cancellation, the origin of which will
be explained at the end of this section.

Taking only neutrino contributions into account, we see from (\ref{n11}) and
(\ref{ccel1}) that the RGE for the vacuum and induced CC are identical
\footnote{%
Of course, this depends on the approximation of constant mass $m$ and
coupling $f$, which we use. At low energies the running of $m$ and $f$ are
negligible.}. Hence the running of the physical CC is governed by the
equation
\begin{equation}
(4\pi )^{2}\frac{d\Lambda _{ph}}{dt}\,=\,-\,4\,\sum_{j}m_{j}^4\,.
\label{newfor}
\end{equation}
Here, as before, we have normalized $t$ such that $t=\ln (\mu /\mu_c)$. Now,
let us make some comment on the cancellation in the one-loop contribution
(\ref{ccel1}).

%%%%%%%%%%%%%%% 1 %%%%%%%%%%%%%%%%
\begin{picture}(120,120)(0,0)
\BCirc(60,40){25} \Text(60,3)[c]{$(a)$}
\end{picture}
%%%%%%%%%%%%%%%%%%%%%%%%%%%%%%%
$\,\,\,\,\,\,\,$
%%%%%%%%%%%%%%% 2 %%%%%%%%%%%%%%%%
\begin{picture}(120,120)(0,0)
\BCirc(60,40){25} \DashLine(85,40)(120,40){4}
\DashLine(0,40)(35,40){4} \Vertex(85,40){2} \Vertex(35,40){2}
\Text(60,3)[c]{$(b)$}
\end{picture}
%%%%%%%%%%%%%%%%%%%%%%%%%%%%%%%
$\,\,\,\,\,\,$ $\,\,\,\,\,\,$
%%%%%%%%%%%%%%% 3 %%%%%%%%%%%%%%%%
\begin{picture}(120,120)(0,0)
\BCirc(60,40){25} \DashLine(0,65)(40,55){4} \Vertex(40,55){2}
\DashLine(0,15)(40,25){4} \Vertex(40,25){2}
\DashLine(80,55)(120,65){4} \Vertex(80,55){2}
\DashLine(80,25)(120,15){4} \Vertex(80,25){2}
\Text(60,3)[c]{$(c)$}
\end{picture}
%%%%%%%%%%%%%%%%%%%%%%%%%%%%%%%%%%%%%%%%%%
%%%%%%%%%%%%%%%%%%%%%%%%%%%%%%%%%%%%%%%%%%%%%%%%%%%%%%%%%%%%%%%%%

\vskip 3mm
\vskip 3mm

\begin{quotation}
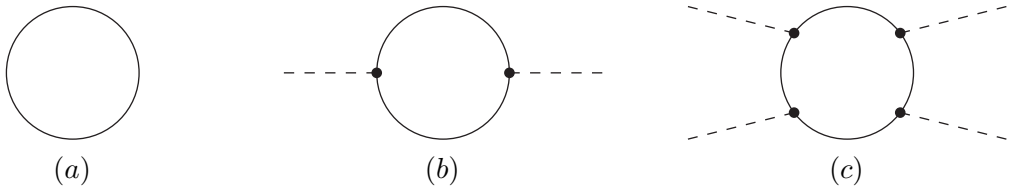
\noindent \textsl{Figure 1.} $\,\,$ {\small \textsl{Three relevant diagrams:
(\textit{a}) The one-loop contributions to the vacuum part are just bubbles
without external lines of matter; (\textit{b}) The one-loop two-point
function contributing to the induced part of the CC. (\textit{c}) The
one-loop four-point function contributing to the induced part.}}
\end{quotation}

%%%%%%%%%%%%%%%%%%%%%%%%%%%%%%%%%%%%%%%%%%%%%%%%%%%%%%%%%%%%%%%%%%%%%%%%%%
\vskip 2mm \vskip 3mm

When investigating the running around $\mu \sim \mu _{c}$, one has to omit
all the diagrams with the closed loops of heavy particles. Then, for the
running of the vacuum CC, one meets only closed neutrino loops without
external tails. In the induced sector, however, there are two sorts of
neutrino diagrams (see \textit{Fig.1}): (\textit{b}) the ones contributing
to the renormalization of the Higgs mass, and (\textit{c}) the ones
contributing to the $\phi ^{4}$-vertex. In the general case and in
dimensional regularization one has
\begin{eqnarray}
\phi _{0} &=&\mu ^{(n-4)/2}\,{Z}_{1}^{1/2}\,\phi \, =\mu ^{(n-4)/2}(1+\frac{1%
}{2}\delta Z_{1})\phi ,\,\,\,\,\,\,\,\,\,  \notag \\
\,\,m_{0}^{2} &=&{Z}_{2}\,m^{2}+\delta {Z}_{3}\,m_{\nu }^{2}\,=(1+\delta
Z_{2})\,m^{2}+\delta {Z}_{3}\,m_{\nu }^{2},\,\,\,\,\,\,\,\,\,  \label{renor}
\\
\,\,f_{0} &=&\mu ^{4-n}\,{Z}_{f}\,\,f=\mu ^{4-n}(1+\delta Z_{f})\,f\,.
\notag
\end{eqnarray}
where $m_{\nu }$ is the neutrino mass, $\delta {Z}_{1},\,\delta {Z}_{2}$ and
$\delta {Z}_{3}$ are divergent one-loop contributions coming from the
diagrams in \textit{\ Fig.1b}, and $\,\delta {Z}_{f}$ comes from the diagram
in \textit{Fig. 1c}. At the low-energy cosmic scale $\mu _{c}$ the heavy
fields do not contribute, so that $\delta {Z}_{2}=0$. But of course at the
Fermi scale a non-trivial $\delta {Z}_{2}$ contribution must be properly
taken into account. Furthermore, at this scale there is an extended list of
one-loop diagrams -- involving the effects from all fermions, Higgs and
gauge bosons of the SM-- from which the general RGE (\ref{n110}) were
derived. However, as we said above, when calculating the $\beta -$function
for $f$ at low energy, we may restrict ourself to the diagrams in Fig.1. As
a result we have the $h_{\nu }^{4}$-order contribution to the $\phi ^{4}$
-vertex from the diagram in \textit{Fig. 1c }plus a $\,fh_{\nu }^{2}$-order
contribution from the tree-level $\phi^{4}$-vertex including the mass
counterterm insertion $\delta {Z}_{3}$ on any of the external legs. This is
the way the ``big'' terms proportional to $fh_{\nu }^{2}$ enter the
calculation. As a consequence, when computing the $\beta $-function for $%
\Lambda _{ind}$ $\ $in Eq.(\ref{ccel1}) \ the $h_{\nu }^{2}$-order
contribution to $m^{2}$ from \textit{Fig. 1b} cancels against the vertex
diagram containing the $\delta {Z}_{3}$ insertion. Since both terms have the
same origin, it is not a real miracle that they cancel out in the RGE for $%
\Lambda_{ind}$. The upshot is that only the $h_{\nu }^{4}$-order
contribution from \textit{Fig. 1c }remains. Since $h_{\nu }^{4}$ is very
small, the running of $\Lambda _{ph}$ has an acceptable range. As for the
higher loop diagrams, it is easy to check that these diagrams come with
extra factors of $h_{\nu }^{2}$, and this renders them much smaller than the
one-loop contributions.

%%%%%%%%%%%%%%%%%%%%%%%%%%%%%%%%%%%%%%%%%%%%%%%%%%%%%%%%%%%%%%%%%%%%
%%%%%%%%%%%%%%%%%%%%%%%%%%%%%%%%%%%%%%%%%%%%%%%%%%%%%%%%%%%%%%%%%%%%
%

\section{The running of CC at higher energies}

%%%%%%%%%%%%%%%%%%%%%%%%%%%%%%%%%%%%%%%%%%%%%%%%%%%%%%%%%%%%%%%%%%%%

In the previous section we have discussed the scaling evolution of the CC at
low energies in the region just above $\mu _{c}$ assuming that only the
lightest massive degrees of freedom\ are active. The study of the heavy
degrees of freedom at higher energies meets several difficulties. Two main
problems are the following: 1) one is the contribution of heavy particles at
the energies near their mass, 2) the other is the ``residual'' effects from
the heavy particles at energies well below their mass. The quantum effects
of the massive particles are,\ in principle, suppressed at low energies \cite
{AC}, so that in the region below the mass of the particle its quantum
effects become smaller, besides, they are not related to the UV divergences.
At this point we need the relation between the IR and the UV regions. The
best procedure to solve 1) would be to extend the Wilson RG for the
quantitative description of the threshold effects, and to solve 2) we would
need a mass-dependent RG formalism. But, since both of these formalisms are
too cumbersome for an investigation at this stage, we will\ first\ of all
tackle the problem by applying the standard \ \ ``sharp cut-off''
approximation within the minimal subtraction (MS) scheme, namely the
contribution of a particle will be taken into account only at the energies
greater than the mass of this particle \footnote{%
It is well-known that the decoupling of heavy particles does not hold in a
mass-independent scheme like the MS, and for this reason they must be
decoupled by hand using the sharp cut-off procedure \cite{Ramond}.}. \
Subsequently, we will roughly estimate the potential modifications of this
approach induced by the heavy particles.

We start by evaluating the successive contributions to the CC up to the
Fermi scale $M_{F}$ within the\ sharp cut-off approximation. The
calculations are performed similarly to the neutrino case at low energy. \
The result is that the $\beta _{\Lambda }$-function in Eq.(\ref{n11}) gets,
in the presence of arbitrary degrees of freedom of spin $J$ and
non-vanishing mass $M_{J},$ a corresponding contribution of the form
\begin{equation}
\beta _{\Lambda }=\,(-1)^{2J}(J+1/2)\,n_{c}\,n_{J}\,\,M_{J}^{4},
\label{betalambda}
\end{equation}
$\,$with $\,(n_{c},n_{1/2})=(3,2)\,$ for quarks, $(1,2)$ for leptons and $%
(n_{c},n_{0,1})=(1,1)$ for scalar and vector fields. The particular case of
the Higgs contribution in Eq.(\ref{n11}) is recovered after including an
extra factor of $4$ from the fact that there are four real scalar fields in
the Higgs doublet of the SM. Notice that this result is consistent with the
expected form $(1/2)\;M_{H}^{4}$ as the physical mass of the Higgs particle
is $M_{H}=\sqrt{2}$ $m$. The values of the CC at different scales, within
our approximation, can be easily computed using the current SM inputs \cite
{Review}. In particular we take $M_{H}=115\,GeV$ and $m_{t}=175\,GeV$. The
numerical results are displayed in Table 1. Notice that the last row gives
the CC at the Fermi scale $M_{F}$. This value follows from integrating the
RGE with the assumption that the masses have their values at the Fermi
scale. From the formulae in Sec. 3 we obtain, after a straightforward
calculation,
\begin{equation}
(4\pi )^{2}\frac{d\Lambda _{ph}}{dt}=\frac{1}{2}M_{H}^{4}+3M_{W}^{4}+\frac{3%
}{2}M_{Z}^{4}-12m_{t}^{4}+\frac{3M_{F}^{4}}{32}\left( 1+\frac{1}{2\cos
^{4}\theta _{W}}\right) g^{4}\,\;\;\;\;(\mu \gtrsim M_{F})  \label{RGEMF}
\end{equation}
We point out that in all cases the contribution from the vacuum and induced
parts is the same to within few percent at most.
%%%%%%%%%%%%%%%%%%%%%%%%%%%%%%%%%%%%%%%%%%%%%%%%%%%%%%%%%%%%%
%%%%%%%%%%%%%%%%%%%%%%%%%%%%%%%%%%%%%%%%%%%%%%%%%%%%%%%%%%%%%
%%%%%%%%%%%%%%%%%%%%%%%%%%%%%%%%%%%%%%%%%%%%%%%%%%%%%%%%%%%%%
%\vskip2mm

\begin{center}
\begin{tabular}{||l|l|l|l||}
\hline\hline
d.o.f. & $m\,(GeV)$ & $\Lambda _{ph}(GeV^{4})$ & $|\Lambda _{ph}|/m^{4}$ \\
\hline
$\nu _{\tau ,\mu }$ & $\approx 10^{-9}$ & $\approx 10^{-47}$ & $\mathcal{O}%
(10^{-11})$ \\ \hline
e & $5\times 10^{-4}$ & $\approx -10^{-37}$ & $\mathcal{O}(10^{-24})$ \\
\hline
u & $5\times 10^{-3}$ & $-3.6\times 10^{-15}$ & $5.8\times 10^{-6}$ \\ \hline
d & $0.01$ & $-3.3\times 10^{-11}$ & $3.3\times 10^{-3}$ \\ \hline
$\mu $ & $0.105$ & $-1.8\times 10^{-9}$ & $1.5\times 10^{-5}$ \\ \hline
s & $0.3$ & $-3.2\times 10^{-6}$ & $4\times 10^{-4}$ \\ \hline
c & $1.5$ & $-9.9\times 10^{-4}$ & $2\times 10^{-4}$ \\ \hline
$\tau $ & $1.78$ & $-0.065$ & $6.7\times 10^{-3}$ \\ \hline
b & $5$ & $-0.33$ & $5.3\times 10^{-4}$ \\ \hline
$W$ & $80$ & $-132$ & $3.2\times 10^{-6}$ \\ \hline
$M_{F}$ & $293$ & $-8.8\times 10^{+7}$ & $0.012$ \\ \hline\hline
\end{tabular}
\end{center}

%%%%%%%%%%%%%%%%%%%%%%%%%%%%%%%%%%%%%%%%%%%%%%%%%%%%%%%%%%%%%
%%%%%%%%%%%%%%%%%%%%%%%%%%%%%%%%%%%%%%%%%%%%%%%%%%%%%%%%%%%%%
\vskip 2mm

\begin{quotation}
%%%%%%%%%%%%%%%%%%%%%%%%%%%%%%%%%%%%%%%%%%%%%%%%%%%%%%%%%%%%%
\noindent {\large \textit{Table 1.}} \textit{The numerical variation of} $%
\Lambda _{ph}$ \textit{\ at different scales }$\mu $\textit{. Each scale is
characterized by the mass }$m$ \textit{of the heaviest, but active, degrees
of freedom (d.o.f.). In the last column, the value of} $\Lambda _{ph}(\mu )$
\textit{is presented also measured in the units of the fourth power of the
natural mass scale} $\mu =m$. \textit{\ Due to the lack of knowledge of the
various neutrino masses, we have given only the order of magnitude of $%
\Lambda _{ph}$ for the second row.}
\end{quotation}

%%%%%%%%%%%%%%%%%%%%%%%%%%%%%%%%%%%%%%%%%%%%%%%%%%%%%%%%%%%%%

\vskip 2mm

We suppose that the heavy couple of neutrinos ($\nu _{\mu },$ $\nu _{\tau }$%
) have masses three orders greater that the masses of the electron and
sterile neutrino (if available). These light neutrinos are assumed with
masses of $\ \mathcal{O}(10^{-3})\,eV$, namely of order of the square root
of the typical mass squared differences obtained in the various neutrino
experiments \cite{Valle}. Since the available data about the neutrino masses
is not exact, their contribution is indicated only as an order of magnitude.
In fact, all of the numbers in this Table are estimates, because of the
reasons mentioned above. Let us make some remarks concerning the values of $%
\Lambda _{ph}$ at different scales. %%%%%%%%%%%%%%%%%%%%%%%%%%%%%%%%%
\vskip 1mm

\textit{First}. The breaking of the fine tuning between induced and vacuum
CC's becomes stronger at higher energies. Even so, it is well under control
because it is highly tamed by the automatic cancellation mechanism in (\ref
{ccel1}). Remarkably, $\Lambda_{ph}$ becomes negative from the heavy
neutrino scale upwards, while its absolute value increases dramatically and
achieves its maximum at the end point of the interval, the Fermi scale.
Notice that at this scale we recover a physical value for the CC around $%
\,10^{8}\,GeV^{4}$, which is of the order of the one obtained from the naive
calculation based on only the (tree-level) induced part, Eq.(\ref{nnn6}).
However, in our framework the value at $\mu =M_{F}$ \ is consistently
derived from a physical CC of order $10^{-47}GeV^{4}$ at the cosmic scale $%
\mu _{c}$. %%%%%%%%%%%%%%
\vskip1mm

\textit{Second}, the dimensionless ratio $\,\Lambda _{ph}/m^{4}\,$ suffers
from ``jumps'' at the different points. Such ``jumps'' occur at the particle
thresholds when the new, heavier, degrees of freedom start to contribute to
the running. Obviously, this is an effect of the sharp cut-off
approximation. In a more precise scheme one has to switch on the
contributions of the heavy particles in a smoother way (e.g. with the aid of
a fully-fledged mass-dependent scheme), and then the scale dependence of the
observable CC would be also smooth. Another possible drawback, although
certainly not inherent to our approach as it is of very general nature, is
the fact that our estimate for the CC at the scale of the light quark masses
may be obscured by non-perturbative effects which are difficult to handle.
Our rough approximation, however, should suffice to conclude that the
relative cosmological constant $\,\Lambda _{ph}/m^{4}\,$, at energies above
the heavy neutrino masses up to the Fermi scale, has a magnitude between $%
10^{-2}$ and $10^{-6}$.

\textit{Third}. One can suppose that there is some (yet unknown) fundamental
principle, according to which the CC is exactly zero at the infinitely small
energy (far IR), that corresponds to the thermodynamical equilibrium by the
end of the evolution of the Universe. It is interesting to verify whether
the change from the non-zero value of CC at present to zero CC at far IR
could be the result of the RG running similar to that we have discussed
above. It is not forbidden at all the existence of the light scalar with the
mass similar to lightes neutrino mass of slightly heavier than the neutrino
mass. In this case the running of the CC could be different from the one
presented in the \textit{Table 1}, in particular the first line or lines of
this \textit{Table} might change the sign. Only experiments can tell us
whether this possibility is real or not. But, such a scalar can not change
the value of CC at far IR, because this scalar has to decouple at the
energies comparable $\mu _{c}$ (simultaneously to neutrino) and can not
affect physics at the scales below $H_{0}$. The only one possibility to
produce the running of CC below $H_{0}$ is to suppose the existence of some
super-light scalar with the mass $m_{sls}\ll H_{0}$. Now, since the observed
value of the CC $\Lambda \approx \mu _{c}^{4}$ is much greater than $%
m_{sls}^{4}$, the only possibility is to have a huge number of copies for
such scalars. On the other hand, there are some serious restrictions for the
number of the super-light scalars, as we shall see in section 7. Hence, the
possibility of having zero CC at far IR and non-zero now, due to the RG,
does not look realistic.

\textit{Fourth}. In the previous considerations, the contributions from the
heavy particles of mass $M$ at energies $\mu \ll M$ are in principle
suppressed by virtue of the decoupling theorem \cite{AC}
\footnote{It is well-known that the decoupling of heavy particles does
not hold in a mass-independent scheme like the MS, and for this reason they
must be decoupled by hand using the sharp cut-off procedure above described
\cite{Ramond} .}. However, one may take a critical point of view and deem
for a moment that their effect could perhaps be not fully negligible in the
context of the CC problem. In this case the modifications of the previous
picture will also depend on the possible choices for the RG scale $\mu $
which, as we have seen in Section 2, is not a completely obvious matter in
the cosmological\ scenario. Whether these heavy mass terms are eventually
relevant or not is not known, but one can at least discuss this possibility
on generic grounds. Under this hypothesis the heavy particle effects
emerging in\ a mass-dependent RG scheme would lead to a new RG equation for $%
\Lambda _{ph}$ in which the \textit{r.h.s.} of eq. (\ref{RGEMF}) ought to be
modified by additional dimension-$4$ terms involving the scale $\mu $
itself, namely terms like $\mu ^{2}M^{2}$. \ This can be guessed from the
fact that in a mass-dependent subtraction scheme\thinspace \cite{Ramond} a
heavy mass $M$ enters the $\beta $-functions through the dimensionless
combination $\mu /M$, so that \ the CC being a dimension-$4$ quantity is
expected to have a $\beta $-function corrected as follows:
\begin{equation}
\beta (m,\frac{\mu }{M})=a\,m^{4}+b\,\left( \frac{\mu }{M}\right) ^{2}M^{4}+%
\mathcal{...}  \label{betaM}
\end{equation}
where $a$,$b$ are some coefficents and the dots stand for terms suppressed
by higher order powers of $\mu /M\ll 1$. Therefore, if we now take e.g. eq. (%
\ref{muH}) as a physical definition of $\mu $ in the gravitational context,
eq.(\ref{betaM}) would lead to a modified RG equation of the generic form
\begin{equation}
(4\pi )^{2}\frac{d\Lambda _{ph}}{dt}=\sum_{i}\,a_{i}m_{i}^{4}+\sum_{j}%
\,b_{j}H^{2}M_{j}^{2}\,\,+...  \label{newRG1}
\end{equation}
\ where $m_{i}$ and $M_{j}$ are the masses of the light and heavy degrees of
freedom respectively, and $\,$the coefficients $b_{i}$ will depend on the
explict mass-dependent computation, but can be expected to be of the order
of the original coefficients $\,a_{i}$ in the mass-independent scheme. The
heaviest masses $M_{i}$ in the SM framework are of order $M_{F}$ and
correspond to the Higgs and electroweak gauge bosons, and the top quark. \ \
In this setting\ the maximum effect from the heavy mass terms is of order $%
\sim H^{2}M_{F}^{2}$. \ Nevertheless if we evaluate the \textit{r.h.s.} of
eq.(\ref{newRG1}) for the present epoch of our universe, there is no
significant contribution other than the residual one because $H_{0}\ll m_{i}$
for any SM particle. So from eq.(\ref{H0}) we infer that the contribution to
$\Lambda _{ph}$ from these terms in the modern epoch is of order \ $%
10^{-82}GeV^{4}\,$at most, i.e. completely negligible. \ From this fact two
relevant conclusions immediately ensue: first, that at present the CC
essentially does not run (a welcome fact if we wish to think of $\Lambda
_{ph}$ as a ``constant''), and second that its value $\sim 10^{-47}GeV^{4}$
was fixed at a much earlier epoch when there was perhaps some active degree
of freedom, like the lightest neutrino (\ref{muc}), and/or when the heavy
mass contributions themselves were of order $10^{-47}GeV^{4}$. In contrast
to these nice results, if we now take the alternative $\mu $ frame (\ref
{muCD}) we find that the contributions $\mu ^{2}M_{i}^{2}$ from the heavy
particles could be much more important even at the modern epoch, rendering a
yield of order $m_{\nu _{1}}^{2}M_{F}^{2}/(4\pi )^{2}\sim 10^{-21}GeV^{4}$
that could distort in a significant way the analysis made in the sharp
cut-off approximation -- unless one arranges for an additional fine tuning
among the various $\mu ^{2}M_{i}^{2}$ contributions \cite{Babic}. As we
shall see in the next section, the choice of $\mu $ can be relevant not only
for the contemporary epoch, but also for the implications in the early
stages of our Universe.

%%%%%%%%%%%%%%%%%%%%    ************   %%%%%%%%%%%%%%%%%%%%%

\section{Implications for the nucleosynthesis}

The first test for the reliability of our effective approach comes from the
primordial nucleosynthesis calculations. The standard version of these
calculations implies that the total energy density $\rho =\rho _{R}+\rho
_{M} $ \ from radiation and matter fields is dominating over the density of
vacuum energy: $\rho \gg \Lambda _{ph}$. In practice, it suffices to verify
that $\rho_{R}\gg \Lambda _{ph}$ because the radiation density is dominant
at the nucleosynthesis epoch. So, we have to check what is the relation
between the CC and the energy density $\rho_{R}$ at the temperature around $%
T_{n}=0.1\,MeV$, which is the most important one for the nucleosynthesis
\cite{Peebles}. If we compare this energy with the electron mass $%
m_{e}\simeq 0.5\,MeV$ and look at the Table 1 above, the plausible
conclusion is that the CC is very small and cannot affect the\ standard
nucleosynthesis results.

However, one has to remind that the nucleosynthesis already starts at the
temperature $10^{10}\,K\simeq 1\,MeV\simeq 2\,m_{e}$. At earlier stages the
entropy is so high that the relevant reactions are suppressed by the high
value of the photon-to-baryon ratio \cite{Peebles}. According to our
previous analysis, that scale of energies is characterized by a fast growth
of the negative CC due to the electron vacuum effects, and above $(5-10)$ $%
MeV$ there is an even greater enhancement due to the light quark
contributions.

Let us now evaluate the energy density of radiation of the Universe at these
temperatures, $\rho _{R}(T)$. It can be obtained from the energy density of
a black body at a temperature $T$. In units where Boltzman's constant is
one, it reads
\begin{equation}
\rho _{R}(T)=\frac{\pi ^{2}}{30}\,g_{\ast }\,T^{4},  \label{rho}
\end{equation}
where $\,g_{\ast }=2$ for photons, and $g_{\ast }=3.36$ after including the
three neutrino species -- if considered \ massless or at least with a mass
much smaller than $T$. Incidentally, for the density of microwave background
photons at the present relic temperature $T_{0}\simeq 2.75K=2.37\times
10^{-4}\,eV$ it gives $\rho _{\mathrm{CMBR}}=2.5\times
10^{-5}\,h_{0}^{-2}\,\rho _{c}^{0}$, i.e. at most one ten-thousandth of the
critical density today ($h_{0}>0.5$) --see eq.(\ref{muc}). However, at very
high energies the density of radiation was dominant. Thus, at the typical
energy of the nucleosynthesis, $\ T_{n}=10^{-4}\,GeV$, we get
\begin{equation}
\rho _{R}(T_{n})\,\simeq \,1.1\cdot 10^{-16}\,GeV^{4}\,.  \label{rhoT}
\end{equation}
The relevant issue at stake now is to check whether the CC at this crucial
epoch of the history of our universe was smaller, larger or of the same
order of magnitud as the radiation energy density. \ If larger, it could
perturb the nucleosynthesis and of course this could not be tolerated. Again
the analysis may depend on the particular\ $\mu $ choices (\ref{muH}) or (%
\ref{muCD}) for the RG scale $\mu $, as well as on the inclusion or not of
the heavy mass terms from eq.(\ref{betaM}). \ Let us forget for the moment
about these terms and start with the definition (\ref{muCD}). Since we now
find ourselves in the radiation dominated epoch, the typical energy density
will be defined by the temperature $T$. So in this approach we may set $\mu
\sim T$ and compare (\ref{rhoT}) with the value of $\Lambda _{ph}(\mu )$
obtained from its scaling evolution with $\mu \sim T$. This can done by
loking at Table 1. Notice that $T$ is of order of $m$ when a particle
species of mass $m$ is active, and so the last column of Table 1 basically
gives the ratio $|\Lambda _{ph}(\mu =M)|/\rho _{R}(T=M)$ up to a factor $%
g_{\ast }\pi ^{2}/15$ of order one. From the Table we realize that the
result (\ref{rhoT}) is about $21$ orders of magnitude bigger than the CC
generated by the ``heavy'' neutrino effects up to the scale $\mu \sim
\,m_{e} $ . Thus, in the framework of our sharp cut-off approximation, the
running of the CC cannot affect the nucleosynthesis. However, the situation
is not that simple, because $T_{n}$ is very close to the electron mass, and
the contribution to the CC from this ''heavy'' particle may become important
at the earlier stages of the nucleosynthesis. In order to see this, let us
derive the density $\rho _{R}$ for the upper energy end of the
nucleosynthesis interval. Using (\ref{rho}) we arrive at the estimate $\rho
_{R}(T=m_{e}=5T_{n})\approx 7\times 10^{-14}\,GeV^{4}\,$ whereas the CC at $%
\mu =m_{e}$ is of order $10^{-37}\,GeV^{4}$. For even higher energies there
is a dramatic enhancement of the CC at $\mu \,\gtrsim m_{u}$ where $\Lambda
_{ph}$ becomes of order $10^{-15}\,GeV^{4}$ whereas $\rho
_{R}(T=m_{u}=50T_{n})\approx 7\times 10^{-10}\,GeV^{4}$. Still, in this case
the CC is five-six orders of magnitude smaller than $\rho _{R}$. So in all
cases $\Lambda _{ph}\ll \rho _{R}$, and this result should not depend on the
sharp cut-off approximation. However, the situation changes dramatically if
the heavy mass terms $\mu ^{2}M_{i}^{2}$ would be present at all in the RGE.
Their presence could be in trouble with nucleosynthesis due to induced
contributions of order $\ T_{n}^{2}M_{F}^{2}/(4\pi )^{2}\sim 10^{-6}GeV^{4}$
which are much larger than (\ref{rhoT}). \ Quite in contrast, the choice (%
\ref{muH}) seems to be completely safe, both with and without heavy mass
terms. \ In fact, from eqs.(\ref{rho}) and (\ref{FL}) Hubble's constant at
the nucleosynthesis time is found to be $H_{n}\sim 10^{-27}GeV$ \ and so $%
H_{n}^{2}M_{F}^{2}/(4\pi )^{2}\sim 10^{-51}GeV^{4}$, \ which is much smaller
than (\ref{rhoT}) and than the present day value of the CC. \ In short, in
the absence of the heavy mass terms $\mu ^{2}M_{i}^{2}$ in the RGE, the
nucleosynthesis cannot be affected by the existence of a renormalization
group induced CC in either of the $\mu $-frames (\ref{muH}) and (\ref{muCD}%
), but if these terms are allowed the nucleosynthesis period could be
jeopardized in the first frame (unless an additional fine tuning is arranged
among the various $T_{n}^{2}M_{i}^{2}$ contributions \cite{Babic}) but it
would remain completely safe in the second frame. \ We \ have thus arrived
from the nucleosynthesis analysis to a similar conclusion as before
regarding the CC value at the contemporary epoch; viz. that the $\mu $
-frame (\ref{muH}) is preferred to the (\ref{muCD}) one for a consistent RG
description of the CC evolution at these two crucial epochs.

%%%%%%%%%%%%%%%%%%%%    ************   %%%%%%%%%%%%%%%%%%%%%
%%%%%%%%%%%%%%%%%%%%    ************   %%%%%%%%%%%%%%%%%%%%%
%

\section{On the running of the gravitational constant}

%%%%%%%%%%%%%%%%%%%%    ************   %%%%%%%%%%%%%%%%%%%%%

Let us consider the running of the gravitational (Newton's) constant, which
can be evaluated in the framework of the algorithm developed for the CC.
From the quantum field theory point of view, the Hilbert-Einstein term
should be introduced into the vacuum action (\ref{7}), because otherwise the
theory is not renormalizable. Then the renormalization condition for the
gravitational constant could be implemented at the scale where it is
measured experimentally, that is at the scale of the Cavendish experiment.

Along with the CC, the induced Hilbert-Einstein term is generated by exactly
the same mechanism as the cosmological term. Disregarding the high
derivative terms, we obtain from Eqs. (\ref{6}) and (\ref{5N}) the action of
induced gravity in the form (\ref{HE}) after replacing $G_{vac}\rightarrow
G_{ind}$ and $\Lambda _{vac}\rightarrow \Lambda _{ind}$, with $G_{ind}$
defined by
\begin{equation}
\frac{1}{16\pi G_{ind}}\,=\,-\,\frac{\xi \,m^{2}}{f}  \label{n6}
\end{equation}
and $\,\Lambda _{ind}\,$ given by (\ref{nnn6}). The physical observable
value of the gravitational constant, $G_{ph}$, obtains from
\begin{equation}
G_{ph}^{-1}=G_{ind}^{-1}(\mu _{c})+G_{vac}^{-1}(\mu _{c})\,.  \label{8}
\end{equation}
When trying to analyze this equation the problem is that the value of $\ $%
the non-minimal parameter $\xi $ is unknown. Indeed, since $G_{ind}$ is
(unlike $G_{ph}$) unobservable, there is no a priori reasonable criterion to
select a value for $\xi $ at any given scale, while the scaling dependence
of $\xi $ is governed by a well-known renormalization group equation (see,
for example, \cite{book}). In the case of the SM we find
\begin{equation}
(4\pi )^{2}\frac{d\xi }{dt}=\left[ \xi -\frac{1}{6}\right] \;\left( 6\,f-%
\frac{9}{2}\;g^{2}-\frac{3}{2}\;g^{\prime
}{}^{2}+2\,\sum_{i}\,N_{i}\,h_{i}^{2}\right) ,\;\;\;\;\;\;\;\;\;\;\;\;\xi
(0)=\xi _{0}  \label{n10}
\end{equation}
where the expression in the parenthesis is the very same one as in the
equation for the mass in (\ref{n110}). We remark, that from the physical
point of view there is no preference at which scale to introduce the initial
data for $\xi (t)$, because this parameter cannot be measured in a direct
way. Some formal arguments can be presented that $\,\xi \approx \frac{1}{6}%
\, $ at high energies \cite{anju} and that it runs very slowly when the
energy decreases \cite{CS}. As we shall see later on, the value of $\xi $ is
not very important for establishing the value of $G_{ph}$ at different
scales.

Now we are in a position to study the scaling dependence for the
gravitational constant. As in the case of the CC, we must consider the
vacuum and induced counterparts independently. The one-loop RGE for the
vacuum gravitational constant can be computed e.g. with the help of the
Schwinger-De Witt technique to extract the divergent part of the one-loop
effective action, and has the form (see, e.g. \cite{frts})
\begin{equation}
(4\pi )^{2}\frac{d}{dt}\,\frac{1}{16\pi G_{vac}}=\,4\,m^{2}\,\left( \xi -%
\frac{1}{6}\right) -\frac{1}{3}\,\sum_{i}N_{i}\,m_{i}^{2}\,,\;\;\;\;\;\;\;\;%
\;\;\;\;\;\;\;\;\;\;G_{vac}(0)=G_{0}\,,  \label{n101}
\end{equation}
where the sum is taken over the spinor fields with the masses $m_{i}$, and $%
\,t=\ln (\mu /\mu _{c})$. The value of $G_{0}$ corresponds to the
renormalization condition at $\mu =\mu _{c\text{ }}$ and must be chosen
according to (\ref{8}). The solution of (\ref{n10}) can be written in the
form:
\begin{equation}
\xi (t)-\frac{1}{6}=\left( \xi _{0}-\frac{1}{6}\right) \;U(t)\,,  \label{n12}
\end{equation}
where $\xi _{0}=\xi (0)$ and $\,U(t)\,$ is defined in (\ref{n113}). Then,
the solution of (\ref{n101}) has the form
\begin{equation}
\frac{1}{16\pi G_{vac}(t)}\,=\,\frac{1}{16\pi G_{0}}+\frac{4}{(4\pi )^{2}}
\,m_{F}^{2}\,\left( \xi _{0}-\frac{1}{6}\right) \;\int_{0}^{t}U^{2}(t)\,dt -%
\frac{1}{3(4\pi )^{2}}\,\sum_{i}\,N_{i}\,\int_{o}^{t}m_{i}^{2}(t)\,dt\,.
\label{n14}
\end{equation}
Thus, the scaling behavior of the parameter $G_{vac}$ is determined, with
accuracy to the integration constant $G_{0}$, by the scaling behavior of the
couplings and masses of the matter fields, and by the initial unknown value $%
\,\,\xi _{0}$.

Consider the induced part. The effective potential of the Higgs field, with
the non-minimal term (\ref{6}), is given by a loop expansion:
\begin{equation}
V_{eff}=V_{cl}+\sum_{n=1}^{\infty }{\hbar }^{n}V^{(n)},  \label{n15}
\end{equation}
where the classical (tree-level) expression
\begin{equation}
V_{cl}=-\frac{1}{2}(m^{2}+\xi R)\,\phi ^{2}+\frac{f}{8}\,\,\phi ^{4}
\label{n16}
\end{equation}
is seen to get an additional contribution from curvature. Since we are
interested only in the running, it is not necessary to account for the
renormalization conditions and one can simply take the renormalization group
improved classical potential. It can be easily obtained from (\ref{n16}) if
all the quantities $\phi ,f,m^{2},\xi $ are replaced by the corresponding
effective charges. The gauge ambiguity related to the anomalous dimension of
the scalar field and consequently with the running of $\phi $ is fixed by
the relation (\ref{5N}). Taking into account (\ref{n6}), we arrive at
\begin{equation}
-\frac{1}{16\pi G_{ind}(t)}=\left[ \frac{1}{6}+\left( \xi _{0}-\frac{1}{6}%
\right) U(t)\right] \;m_{F}^{2}\;U(t)\;f^{-1}(t)\,,  \label{n17d}
\end{equation}
where $f(t)$ is solution of Eq. (\ref{n110}). The formulas above give the
scaling dependence for the induced gravitational constant, which is
completely determined by the running of $m^{2}(t)$, $f(t)$ and $\xi (t)$.
Eq. (\ref{8}) enables one to establish the scale dependence of the physical
gravitational constant:
\begin{equation*}
-\frac{1}{16\pi G_{ph}(t)}=-\frac{1}{16\pi G_{0}}+\left[ \frac{1}{6}+\left(
\xi _{0}-\frac{1}{6}\right) U(t)\right] \;\frac{m_{F}^{2}}{f(t)}\,U(t)\,-
\end{equation*}
\begin{equation}
-\frac{\,4\,m_{F}^{2}\,}{(4\pi )^{2}}\left( \xi _{0}-\frac{1}{6}\right)
\int_{0}^{t}U^{2}(t)dt\,+\,\frac{1}{3(4\pi )^{2}}\,\sum_{i}\,N_{i}\,
\int_{o}^{t}m_{i}^{2}(t)\,dt\,.  \label{n17}
\end{equation}

From the last formula follows, that the scaling dependence of the inverse
gravitational constant (deviation of its value from $\,G_{0}$) is
proportional to $\,M_{F}^{2}\,\sim \,10^{5}\,GeV\,^{2}$ whereas the
observable value of $G_{ph}^{-1}$ is $M_{Pl}^{2}\sim 10^{38}\,GeV^{2}$.
Hence, the only one chance to have relevant running of $G_{ph}$ is to
consider the theory with huge $\xi $ comparable with $10^{33}$. It is easy
to see that this can be inconsistent with our general supposition in Eq.(\ref
{n16}) that $\xi R$ is small as compared to $m^{2}$ in the Fermi epoch. We
use this fact when take the flat-space formula (\ref{2a}) for the SSB,
despite our potential contains a non-minimal term (\ref{6}). In order to
justify this, one has to remind that the values of Ricci tensor and
energy-momentum tensor are linked by Einstein equations (\ref{Einstein} The
typical value of the components of the physical $T_{\mu \nu }$ is $\mu^{4}$
where $\mu $ is the scale at the corresponding epoch. At the Fermi epoch, $%
\mu \sim m\sim 100\,GeV$, the typical value of the components of $R_{\mu \nu
}$ is of order $R\sim 8\pi GT_{\mu }^{\mu }\sim 8\pi\,m^{4}/M_{P}^{2}$, and
so indeed our approximation (\ref{2a}) works well only if
\begin{equation}
|\xi |<<\frac{1}{8\pi }\left( \frac{M_{P}}{m}\right) ^{2}\sim 10^{33}\,.
\label{xi}
\end{equation}
On the other hand, since the scale dependence of $\xi $ (\ref{n12}) is not
strong \cite{CS}, $\xi $ must be close to $\,1/6\,$ also at the present
epoch. If being of the order $10^{33}$, $\xi $ should manifest itself in the
low energy phenomena, and since this is not the case we will not consider
this possibility. For values of $\xi $ satisfying (\ref{xi}) the running of $%
G_{ph}$ is negligible so that the exact value of $\xi $ is not important for
the definition of the gravitational constant.

\ Finally, we remark that the negligible running of $G_{ph}$ that we have
found in our framework, is very much welcome in order not to disturb
primordial nucleosynthesis, which in fact can only tolerate few percent
deviations of $G_{ph}$ with respect today's value \cite{Rubakov}. This
feature, together with the already proven relative smallness of the vacuum
energy as compared to the energy density of matter at that epoch, is quite
rewarding for the physical consistency of our approach.

\ As for the potential heavy mass corrections to the
previous analysis, similar considerations can be made here in complete
analogy with the CC case. \ However, in the present case the dimensionality
of\ $G_{ph}^{-1}$ enforces the additional terms to be of the form $\left(
\mu /M\right) ^{2n}M^{2}$ $(n=1,2,...).$ \ Hence the first, and most
important, correction is just $\mu ^{2\text{ }}$(independent of $\ $the
heavy mass $M$ ). \ Then, it is obvious that the presence of these terms
would be inncocuous because $\mu ^{2}\ll G_{ph}^{-1}$ for both $\mu$-frames (%
\ref{muH}) and (\ref{muCD}).

%%%%%%%%%%%%%%%%%%%%    ************   %%%%%%%%%%%%%%%%%%%%%
%%%%%%%%%%%%%%%%%%%%    ************   %%%%%%%%%%%%%%%%%%%%%

\section{Cosmological constant and inflation}

%%%%%%%%%%%%%%%%%%%%    ************   %%%%%%%%%%%%%%%%%%%%%

It is well known that many problems of standard cosmology can be solved via
the inflationary paradigm \cite{Guth}. Then, it is common wisdom to suppose
that some kind of induced CC was responsible for the inflation. But, in the
phenomenologically acceptable scenarios, one can achieve the successful
inflation through the inflaton models \cite{Guth}, while the origin of the
inflaton remains unclear. There is an alternative scenario of inflation,
which originates from the quantum vacuum effects of matter fields \cite
{star,vile,anju}. These effects are related to the renormalization of the
higher derivative sector in (\ref{7}) and do not suppose the crucial role
for the cosmological constant. However, it is interesting to check, how the
CC affects the inflationary solution.

In order to arrive at the anomaly-induced effective action, one has to start
from the quantum theory of matter fields which are conformally coupled to
gravity. In the framework of the cosmon model \cite{PSW} one can admit the
existence of the masses, but in any case one has to request the $\xi =1/6$
condition \cite{shocom}. Then the high derivative sector of the vacuum
action (\ref{7}) can be reduced to
\begin{equation*}
S_{vac}=\int d^{4}x\sqrt{-g}\,\Big(\,l_{1}C^{2}+l_{2}E+l_{3}{\square }R\,%
\Big)\,,  \label{vacuum}
\end{equation*}
where, $l_{1,2,3}$ are some parameters, $C^{2}$ is the square of the Weyl
tensor and $E$ is the integrand of the Gauss-Bonnet topological invariant.
the renormalization of the action (\ref{vacuum}) leads to anomaly
\begin{equation}
T=<T_{\mu }^{\mu }>\,=\,-\,(wC^{2}+bE+c{\square }R)\,,  \label{anomaly}
\end{equation}
where $\,w,\,b,\,c\,$ are the $\beta $-functions for the parameters $%
\,\,l_{1},\,l_{2},\,l_{3}\,\,$ \footnote{%
The contribution of the Weyl spinor is half of that for the Dirac spinor and
the contribution of complex scalar is twice the one of the real scalar.}
\begin{equation}
w=\frac{N_{0}+6N_{1/2}+12N_{1}}{120\cdot (4\pi )^{2}}\,\,,\,\,\,\,\,\,\,b=-\,%
\frac{N_{0}+11N_{1/2}+62N_{1}}{360\cdot (4\pi )^{2}}\,\,,\,\,\,\,\,\,\,c=%
\frac{N_{0}+6N_{1/2}-18N_{1}}{180\cdot (4\pi )^{2}}\,.  \label{abc}
\end{equation}
The anomaly-induced action, for the massless fields has the following form
\cite{rei}:
\begin{equation}
{\bar{\Gamma}}=S_{c}[{\bar{g}}_{\mu \nu }]+\int d^{4}x\sqrt{-{\bar{g}}}\,\{w{%
\bar{C}}^{2}+b({\bar{E}}-\frac{2}{3}{\bar{\nabla}}^{2}{\bar{R}})+2b{\bar{%
\Delta}}\sigma +d{\bar{F}}^{2}\,,  \label{massless}
\end{equation}
where we have included the contribution of the vector fields. Here the
original metric is decomposed according to ${\bar{g}}_{\mu \nu }=g_{\mu \nu
}\,e^{2\sigma }$ and $S_{c}[{\bar{g}}_{\mu \nu }]$ is the conformal
invariant part of the quantum contribution to the effective action, which is
an integration constant for the solution (\ref{massless}). Adding up (\ref
{massless}) with the Hilbert-Einstein term (\ref{HE}) and performing the
variation of the total action with respect to $a(t)=e^{\sigma (t)}$ ($t$ is
the physical time) one obtains the corresponding equation of motion \cite
{anju}:
\begin{equation}
E[a]=a^{2}{\overset{....}{a}}+3a{\overset{.}{a}}{\overset{...}{a}}+\left( 5-%
\frac{4b}{c}\right) {\overset{.}{a}}^{2}{\overset{..}{a}}+a{\overset{..}{a}}%
^{2}-\frac{2M_{P}^{2}}{c}\,a\,\left( a{\overset{..}{a}}+{\overset{.}{a}}%
^{2}\right) =0  \label{equation}
\end{equation}
This equation leads to the exponential solution of Starobinsky \cite{star}
\begin{equation}
a(t)=e^{H_{P}t}\,,\,\,\,\,\,\,\,\,\,\,\,\,\,\,\,H_{P}=\frac{M_{P}}{\sqrt{-b}}%
\,,  \label{exponential}
\end{equation}
which is stable for the particle content $N_{1}$ vectors, $N_{1/2}$ spinors
and $N_{0}$ scalars satisfying the inequality
\begin{equation}
N_{1}\,<\,\frac{1}{3}\,N_{1/2}\,+\,\frac{1}{18}\,N_{0}\,.  \label{const}
\end{equation}
The relation (\ref{const}) allows one to distinguish between very different
scenarios of non-stable inflation \cite{star,vile} and the stable one \cite
{anju}. Our purpose here is to verify how the solution (\ref{exponential})
and the stability condition (\ref{const}) look in the presence of the
cosmological constant. Let us formulate a more general question: how will
Eq. (\ref{equation}) modify in the presence of some source term (matter or
CC) which satisfies some equation of state? It is easy to see that any
matter content produces an energy density which rapidly decrease with the
growth of $a(t)$. Substituting such a source term into (\ref{equation}), and
making numerical analysis, we find that the presence of such source, even
with initial energy density of the Planck order of magnitude, does not
perturb in any essential way the stable solution (\ref{exponential})
\footnote{%
Authors are grateful to Ana Pelinson for discussion of this point.}. This
fact indicates, that the ''equation of state'' for the high derivative terms
in (\ref{massless}) is such that the corresponding energy density does not
change with $a(t)$. Since the same property is shared by the CC, one can
expect that the CC will be the only possible source of modifications for the
stable inflationary solution. Let us check this. The equation with the CC
source has the form \footnote{%
In this equation we disregard the scale dependence of the CC. This is
justified, because this dependence has logarithmic structure and the result
can not change too much if we use another approximation for the $\Lambda
_{ph}(a)$.}:
\begin{equation}
E[a]=-\frac{2\Lambda _{ph}}{3c}\,a^{3}\,,  \label{equation1}
\end{equation}
where the operator $E[a]$ has been defined in (\ref{equation}). Taking into
account that the logarithmic dependence is weak, for the sake of simplicity
we consider that $\Lambda _{ph}$ is constant with respect to $a$.
Substituting the exponential function $a(t)=e^{Ht}$ into (\ref{equation1}),
we arrive at two solutions for the induced gravity with the CC (see also
\cite{odintsov} where similar solutions have been found):
\begin{equation}
H_{1}^{2}=\frac{1}{2}\,H_{P}^{2}+\frac{1}{2}\,\sqrt{H_{P}^{2}+\frac{2}{3b}%
\Lambda _{ph}}\,,\,\,\,\,\,\,\,\,\,\,\,\,\,\,\,\,H_{2}^{2}=\frac{1}{2}%
\,H_{P}^{2}-\frac{1}{2}\,\sqrt{H_{P}^{2}+\frac{2}{3b}\Lambda _{ph}}\,.
\label{new solutions}
\end{equation}
In the case of the sufficiently small CC he first solution is close to the
inflationary Starobinsky solution (\ref{exponential}). At the inflationary
epoch, the CC could include the contributions from the running due to the
massive constituents of GUT, and also due to the phase transitions at varios
scales. But, for the consistency of the semiclassical approach \cite{anju}
one has to suppose that all the particle masses are much smaller than $M_{P}$%
. Therefore the first solution in (\ref{equation}) manifests just a small
difference with $H_{P}$, such that we always have $H_{P}^{4}\propto
M_{P}^{4} $. The sign of $\Lambda _{ph}$ does not play much role in this
context.

The second solution is also interesting, because it substitutes the flat
static solution $\,a(t)\equiv a_0=const$ of Eq. (\ref{equation}). This
solution applies at the very low energy. For example, at the cosmic scale
the value of $\Lambda_{ph}$ is positive and of the order of $\rho_M^0$. Let
us consider this solution in details, disregarding the effect of matter
density (which will be considered elsewhere). In the presence of the CC this
flat solution becomes some very slow inflation, such that the effect of the
anomaly-induced term is not seen. In order to check the last statement, we
remind that the CC is very small $\Lambda_{ph}\ll M_P^4$. Expanding the
second solution $H_2$ up to the first order we find that the effective
exponential behavior is governed by $H_{eff}=\Lambda_{ph} /(6 M_P^4)$. It is
easy to check that this is exactly the value which emerges in the Einstein
theory with the CC but without higher derivative induced terms (\ref
{massless}). Of course, the inflationary solution exists only for the
positive $\Lambda_{ph}$.

Let us discuss the stability of the two solutions (\ref{new solutions})
under the perturbations of the conformal factor of the metric $a(t)$
\footnote{%
The detailed investigation of the stability problems will be given elsewhere
\cite{stab}.}. For this purpose, one has to expand, in both cases,
\begin{equation}
H(t)\,=\,H_{1/2}+x(t)  \label{perturb}
\end{equation}
and, after linearization, investigate the asymptotic behavior of the
perturbation $x(t)$. It is easy to see, that for the physically interesting
case $\Lambda _{ph}\ll M_{P}^{4}$ the stability condition for the first
solution $a_{1}(t)=e^{H_{1}t}$ is nothing but (\ref{const}). This is quite
natural, because the cosmological constant does not play much role in this
solution. With the opposite sign of the inequality in (\ref{const}), the
second solution is stable, exactly as its flat counterpart in the zero CC
case.

The last observation concerns the possibility to have many copies of a very
light scalars which we have discussed by the end of section 4. Since the
presence of the CC does not change the stability condition (\ref{const}),
the existence of too many scalars would unavoidably produce a very fast
inflation. Since today only the photon (among the known particles) remains
active, the condition (\ref{const}) predicts that the present-day Universe
must live close to the non-inflationary stable state. But, if we suppose
that there are more than 18 super-light scalars, they would not decouple
below $\mu _{c}$, the Universe would be inflating at the corresponding scale
and this could break the nucleosynthesis. Thus, these scalars are ruled out
by observations.

\ Finally, as in the \ $G_{ph}^{-1}$ \ case, we point out that the potential
heavy mass contributions in the RGE are also of residual nature here because
$\mu ^{2}\ll H_{P}^{2}$ for both $\mu $-frames (\ref{muH}) and (\ref{muCD}),
so that they cannot alter in any significant way the conclusions derived
from eqs. (\ref{new solutions}).

%%%%%%%%%%%%%%%%%%%%    ************   %%%%%%%%%%%%%%%%%%%%%
%

\section{Conclusions}

\ We have considered several new aspects of the CC problem
using the RG method. The main difference between the CC and the other SM
parameters is that the input data on the CC are only known at the cosmic
scale $\mu _{c}$ defined in (\ref{muc}), and therefore the CC can only be
renormalized at this scale, whereas the other parameters are renormalized at
the Fermi scale $M_{F}\gg \mu _{c}$. This hints at the
necessity for the unnaturally great precision of the vacuum CC value at $\mu
=\mu _{c}$. If we take the cancellation of the vacuum and
induced CC's at $\mu =\mu _{c}$ as an experimental fact, the standard
quantum field theory framework predicts the running of the CC above the
scale $\mu _{c}$, and also restricts the kind of new (massive) particles
potentially existing below that scale.

Our investigation shows that the running of the CC could take place as the
expanding universe progressively cooled from energies comparable to the
Fermi scale (or above) down to the modern epoch, but in actual fact the
running itself got stuck at the scale $\mu _{c}$, which is comparable to the
lightest neutrino mass, $m_{\nu _{1}}\sim 0.001\,eV$. This scale
is many orders of magnitude greater than the present-day
cosmic scale, defined by the value of Hubble's constant $H_{0}\sim
10^{-33}\,eV$. Still, the range of the running of the CC is of the same
order of magnitude ($\Lambda _{ph}\sim 10^{-47}\,GeV^{4}$) as the energy
density of matter in the present Universe. The existence of ultralight
(sub-gravitationally coupled) degrees of freedom below the scale $H_{0}$ is
not, in principle, excluded. However, they cannot be responsible for a
vanishing CC in the far IR through the RG mechanism. Finally, we may get a
better understanding of the aforementioned ``cosmic coincidences'' as
follows. On the one hand $\rho _{M}^{0}\sim \rho _{c}^{0}$, and on the other
it turns out that $m_{\nu _{1}}^{4}\sim \rho _{c}^{0}$. In our framework
these two facts can be reconciled with the third coincidence, $\rho
_{M}^{0}\sim \Lambda _{ph}$, because the CC must be generated through the RG
running, and naturally acquires a value comparable to the fourth power of
the mass of the lightest (massive) degree of freedom: $\Lambda _{ph}\sim
m_{\nu _{1}}^{4}$. Whether the neutrino is ultimately responsible for this
value cannot be established at the moment, but the correct\ (positive) sign
of the CC may require the existence of a light scalar of comparable mass.
Recall that values of the CC significantly greater than $m_{\nu _{1}}^{4}$
might contradict well-known anthropic considerations \cite{weinRMP,weinDM}.

Another important conclusion of our semiclassical RG analysis of the CC
problem is the following: Of the two possible choices (\ref{muH}) and (\ref
{muCD}) for the scale $\mu $, definition (\ref{muH}) seems to be the most
appropriate one for a consistent description. First, (\ref{muH}) sets the
natural scale of the graviton energy as read off Einstein equations applied
to the FLRW metric. Second, if heavy mass corrections to the sharp cut-off
RG analysis need eventually be included, definition (\ref{muH}) is the only
one that allows to introduce the appropriate modifications in the RGE in a
way that is naturally compatible (i.e. without any
additional fine tuning) with both the value of the CC at present and the
value of the CC at the epoch of the nucleosynthesis.
Moreover, the existence and adequacy of the $\mu $-frame
(\ref{muH}) shows that for the semiclassical RG analysis of the
cosmological parameters we must actually distinguish between two
energy scales: One scale $\mu $ refers to the typical energy of
the external gravitons, and another scale $\mu^{\prime}$ refers to
the typical energy of the external matter particles -- the latter
being closer to the frame (\ref{muCD}). While the former is
relevant for the running of the cosmological parameters the
latter characterizes the scaling evolution of the SM couplings
and masses. Since the two scales $\mu $ and $\mu^{\prime}$
typically run on vastly different regimes ($\mu \sim H$ \ and
$\mu^{\prime}\sim M_{F}$), we are unavoidably confronted with the
necessity of an apparent large fine tuning when we add up the
induced and vacuum contributions to the CC in order to build up
the physical value defined in eq.(\ref{lambda_phys}): $\Lambda
_{ph}=\Lambda _{vac}(\mu =\mu _{c})+\Lambda
_{ind}(\mu^{\prime}=\mu _{c})$.
The fine tuning is prompted by the fact that the CC renormalization point $%
\mu _{c}$ lies in the vecinity where $\mu $ typically runs while
it is exceedingly away from the typical range of $\mu^{\prime}$.
The two scales can never be close because the graviton energy is
always suppressed by a factor $M_P$, more precisely:
$\mu/\mu^{\prime}\sim\mu^{\prime}/M_P\sim M_F/M_P$.

Similarly to the CC, there is a running of the gravitational constant. We
have proven that the scope of this effect is small and that it can not spoil
primordial nucleosynthesis. Furthermore, we have found that the CC does not
distort the virtues of the anomaly-induced inflation. At high energies the
presence of the CC does not modify, in any essential way, neither the
velocity of inflation nor the condition of stability (\ref{const}). Finally,
at low energies the anomaly-induced terms do not change the observed value
of the CC.

%%%%%%%%%%%%%%%%%%%%    ************   %%%%%%%%%%%%%%%%%%%%%
%
\vskip3mm \noindent \textbf{Acknowledgments.} I.L.Sh. is grateful
to M. Asorey, J.C. Fabris, L.P. Grishchuk, V.N. Lukash and A.M.
Pelinson for helpful discussions and is indebted to CNPq
(Brazil), Grup de F\ii sica Te\`{o}rica/IFAE at the Univ.
Aut\`{o}noma de Barcelona, and the Dep. E.C.M. at the Univ. de
Barcelona for support. The work of J.S. has been supported in
part by CICYT under project No. AEN99-0766, by the Generalitat de
Catalunya (project BE 2000) and by CNPq (Brasil); J.S. thanks A.
Goobar and J.W.F. Valle for useful conversations.

%\newpage
%%%%%%%%%%%%%%%%%%%%%%%%%%%%%%%%%%%%%%%%%%%%%%%%%%%%%%%
%

\end{document}